\documentclass[twocolumn,showpacs,preprintnumbers,amsmath,amssymb,superscriptaddress]{revtex4}
\pdfoutput=1

\usepackage{float,graphicx,subfigure}
\usepackage{dcolumn,multirow}
\usepackage{bm}
\usepackage{psfrag}

\begin{document}
\title{Effect of next-nearest neighbor interactions on the dynamic order parameter of the Kinetic Ising model in an oscillating field}
\author{William D. Baez}
\affiliation{Department of Chemistry and Physics, Augusta State University, Augusta, GA 30904}
\author{Trinanjan Datta} 
\affiliation{Department of Chemistry and Physics, Augusta State University, Augusta, GA 30904}
\email{tdatta@aug.edu}
\date{\today}
 
\begin{abstract}
We study the effects of next-nearest neighbor (NNN) interactions in the two-dimensional ferromagnetic kinetic Ising model exposed to an oscillating field. By tuning the interaction ratio ($p=J_{NNN}/J_{NN}$) of the NNN ($J_{NNN}$) to the nearest-neighbor (NN) interaction ($J_{NN}$) we find that the model undergoes a transition from a  regime in which the dynamic order parameter $Q$ is equal to zero to a phase in which $Q$ is not equal to zero. From our studies we conclude that the model can exhibit an \emph{interaction induced transition from a deterministic to a stochastic state}. Furthermore, we demonstrate that the systems' metastable lifetime is sensitive not only to the lattice size, external field amplitude, and temperature (as found in earlier studies) but also to additional interactions present in the system. 
\end{abstract}

\maketitle

\section{Introduction}
The two-dimensional (2D) ferromagnetic nearest-neighbor kinetic Ising (NNKI) model in an oscillating field has been used extensively to study non-equilibrium (NEQ) properties ~\cite{Chakrabarti:RevModPhys.71.847,Sides:PhysRevLett.81.834}. The presence of an oscillatory field introduces an explicit time dependence in the Hamiltonian. This causes the system to exhibit a hysteretic response. Below the critical temperature ($T_c$) the system exhibits a dynamic phase transition (DPT) or stochastic resonance (SR) ~\cite{Chakrabarti:RevModPhys.71.847,Sides:PhysRevLett.81.834,Sides:PhysRevE.59.2710,Korniss:PhysRevE.63.016120,Korniss:PhysRevE.66.056127} based upon the strength of the field amplitude ($h_0$), frequency ($f$), temperature ($T$) of the system, and the lattice size ($L$). These parameters dictate whether the system will be in a deterministic or a stochastic regime. These regimes are further subdivided into one of the four possible regions: strong field (SF) and multi-droplet (MD) for deterministic, single-droplet (SD) and co-existence (CE) for stochastic. The SF and the MD regions exist for strong fields/large system sizes and the SD exists for weak fields/small system sizes ~\cite{Sides:PhysRevLett.81.834}. The MD and SD regions have been studied exhaustively for metastability mechanisms, finite-size scaling effects, DPT, SR, hysteresis exponents, universality, critical exponents, and effect of square-wave oscillating external field with a soft Glauber dynamics ~\cite{Sides:PhysRevE.57.6512,Sides:PhysRevE.59.2710,Korniss:PhysRevE.63.016120,Korniss:PhysRevE.66.056127,buenda:051108,PhysRevLett.92.015701}. 

A measure of the DPT is given by the dynamic order parameter, $Q$, which is the period-averaged magnetization. The DPT occurs between an ordered dynamic phase with $\langle|Q|\rangle \neq 0$ and a disordered dynamic phase with $\langle|Q|\rangle = 0$ ~\cite{Sides:PhysRevLett.81.834} (only in the MD region).  The DPT in the NNKI model is of second order ~\cite{Sides:PhysRevLett.81.834,Sides:PhysRevE.59.2710,Korniss:PhysRevE.63.016120}. Recently, evidence for a dynamic phase transition has been investigated experimentally in [Co/Pt]$_3$ magnetic multilayers ~\cite{robb134422}. Furthermore, Monte Carlo simulations and mean-field studies of hysteresis in the NNKI model show that the loop area undergoes a transition from a symmetric shape to an asymmetric shape or vice-versa ~\cite{Chakrabarti:RevModPhys.71.847,Liu:PhysRevB.70.132403}. The breakdown in the shape of the hysteresis loop has also  been attributed to a DPT with a \emph{spontaneously broken symmetric phase} ~\cite{Chakrabarti:RevModPhys.71.847}. The scaling relations for the hysteresis loop area have been measured in ultrathin and thin ferromagnetic film systems such as Fe/Au(001) ~\cite{PhysRevLett.70.2336}, Fe$_{20}$Ni$_{80}$ ~\cite{PhysRevB.60.11906}, and Co/Cu(001) ~\cite{PhysRevB.52.14911}.  In the SR region the magnetization switches through random nulcleation of a single droplet of spins aligned with the applied field. This region has been studied for stochastic hysteresis using time-dependent nucleation theory and a variable rate Markov processes ~\cite{Sides:PhysRevE.57.6512}.

In this paper we extend the NNKI model to include next-nearest neighbor (NNN) feromagnetic interactions. We term this model the next-nearest neighbor kinetic Ising (NNNKI) model (see Eq.~\ref{eq:NNNKI}). The  motivation behind this investigation is to study the effects of additional (NNN) interactions in the system. In Sec.~\ref{sec:modelmethod} we state the NNNKI model and describe the Monte Carlo method used. In Sec.~\ref{sec:results} we present our results on the NNNKI model. In Sec.~\ref{sec:disconcl} we discuss the results and state the main conclusions of our paper. 

\section{Model and Method}\label{sec:modelmethod}
The model used in this study is the kinetic NNN Ising ferromagnet on a square lattice with periodic boundary conditions. The NNNKI Hamiltonian is given by \begin{eqnarray}\label{eq:NNNKI}
H=-J_{nn}\sum_{\langle i,j\rangle}S_{i}S_{j}-J_{nnn}\sum_{[i,j]}S_{i}S_{j}-h_{o}\sin(2\pi f t)\sum_{i}S_{i},
\end{eqnarray} where $S_{i}$ is the $i$th spin and can have values of $S_{i}=\pm$ 1, $J_{nn}$ is the NN coupling, $J_{nnn}$ is the NNN coupling, $h_o$ is the external field amplitude, and $f$ is the frequency of the external field. The sums $\sum_{\langle i,j\rangle}$ and $\sum_{[i,j]}$ run over all NN and NNN pairs, respectively. Both the couplings are ferromagnetic, $J_{nn}>0$ and $J_{nnn}>0$. The ratio of the couplings is defined as $p=J_{nnn}/J_{nn}$. The spin-flip dynamics used is the Metropolis algorithm with the Monte Carlo step per spin (MCSS) as the unit time step ~\cite{LandauBinderbook}. The system is allowed to be in contact with a heat bath at temperature T, and each attempted spin flip from $S_{i}\rightarrow -S_{i}$ is accepted with the probability $W(S_{i}\rightarrow -S_{i})=\exp(-\beta \Delta$ E$_{i}$). Here $\Delta E_{i}$ is the change in energy of the system that would result if the spin flip were accepted and $\beta=1/k_{B}T$ where the Boltzmann constant $k_{B}$ is set equal to one. Using the above Hamiltonian (Eq.~\ref{eq:NNNKI}) and the Monte Carlo method we compute the dynamic order parameter $Q$
\begin{equation}
{\label{eq:Qform}Q=\frac{\omega}{2\pi}\oint m(t) dt},
\end{equation} 
where $m(t)$ is the time dependent magnetization per unit site. In the next two sections we study the physics of the NNNKI model. 
\begin{widetext}
\begin{figure}[htb!]
\centering
\mbox{\subfigure{\includegraphics[width=3in]{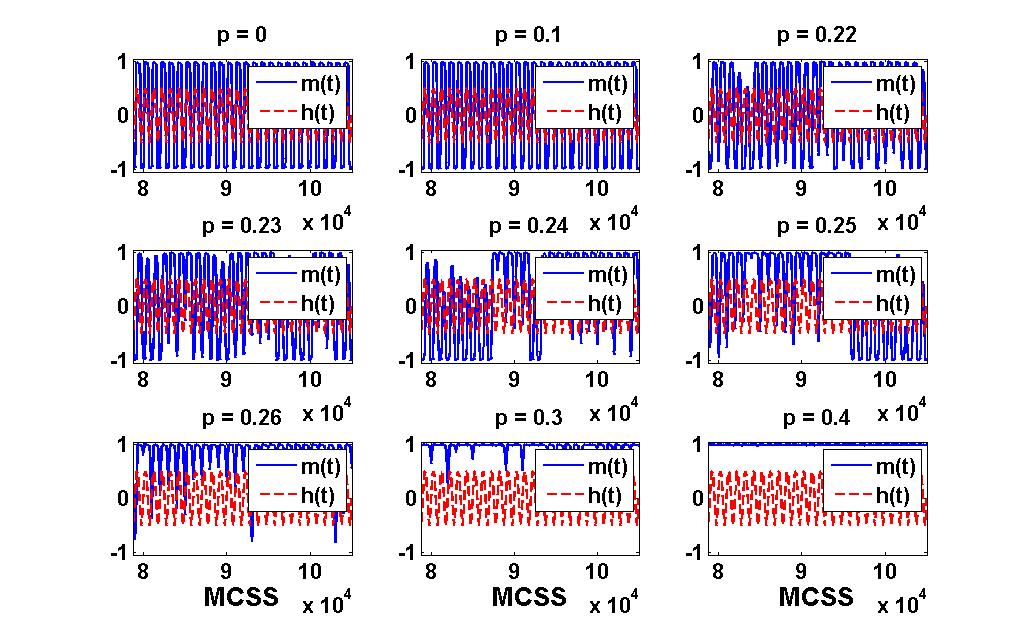}\label{subfig:magtimeseries}}
\subfigure{\includegraphics[width=3in]{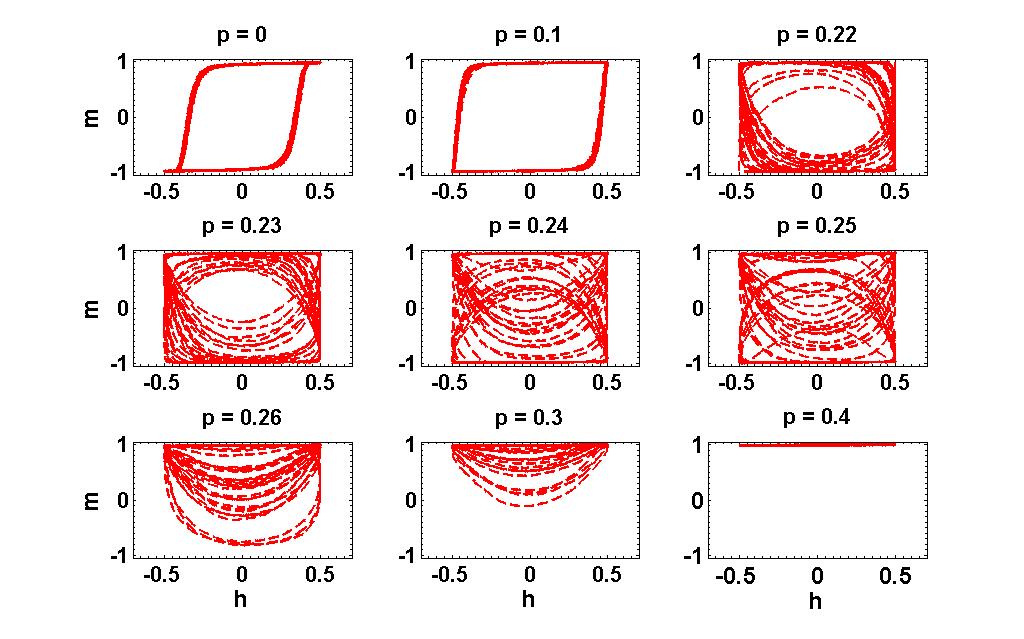}\label{subfig:hyst}}}
\caption{For $L=128$, $h_{o}=0.5J_{nn}$ (where $J_{nn}=1$), $f=10^{-3}$, and $T=0.8T_{c}^{NN}$ we have (a) Magnetization time series data demonstrating the deterministic to stochastic transition for interaction ratio p = 0 - 0.4. The solid line represents the systems magnetization and the dashed line is the external field. (b) Spontaneous symmetry breaking of the hysteresis loops for the same set of interaction ratios as in part (a).}
\label{fig:1}
\end{figure}
\end{widetext}
\begin{figure}[htb!]
\centering
{\includegraphics[width=3.5in]{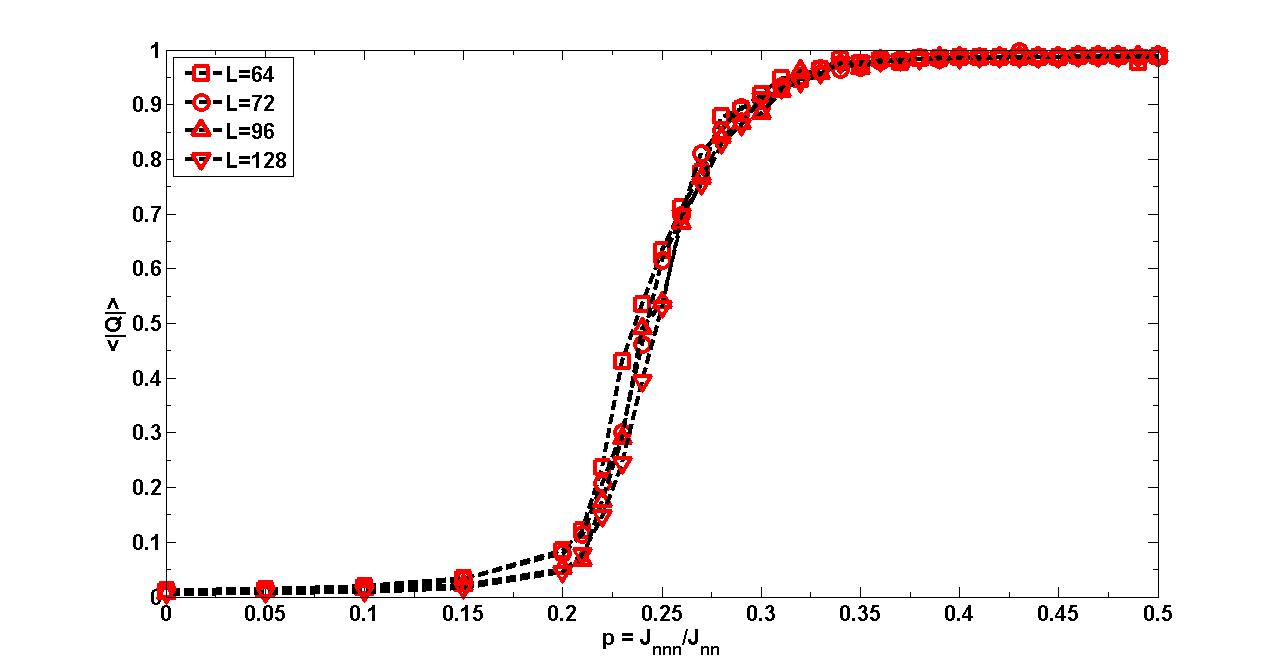}}
\caption{Tracking the dynamic order parameter, $\langle |Q|\rangle$, as the interaction ratio, $p=J_{nnn}/J_{nn}$, is changed. The value of $\langle |Q|\rangle$ changes from a zero to a non-zero number. The lattice sizes are $L$=64, 72, 96, and 128. The computations were done at $h_{o}=0.5J_{nn}$ (where $J_{nn}=1$), $f=10^{-3}$, and $T=0.8T_{c}^{NN}$. Each run is of length 100,000 MCSS. The dashed lines are a guide to the eye.}\label{fig:2}
\end{figure}
\section{Results}\label{sec:results}
\subsection{Interactions and their effects\label{sec:pq}}
In all the previous studies involving the NNKI model the frequency, magnetic field or temperature was used as the tuning parameter to demonstrate a transition between the stochastic to deterministic region. This allowed the study of SR or DPT based upon the system size and field amplitude. Our goal in this section is to demonstrate that the deterministic to stochastic transition can also be achieved by \emph{tuning the interactions}. In our computation we choose a set of ($h_{o},f,T$) values so that the system is in the deterministic MD region in the NN model. We take $h_{o}=0.5J_{nn}$, $f=10^{-3}$, and $T=0.8T_{c}^{NN}$. A choice of ($h_{o}$,f,T) so that the system is in stochastic region is also equally valid. We then change the values of interaction ratio, $p$, and compute the magnetization time series data for several values of $p$ (see Fig.~\ref{subfig:magtimeseries}). In our calculations we take $p$ to range from 0 to 0.4. This range of parameters is sufficient to highlight the physics of the problem. From the magnetization time series data we see that the system makes a transition from a deterministic to a stochastic region. For the same values of $p$ we plot the hysteresis loops which show the spontaneous symmetry breaking (see Fig.~\ref{subfig:hyst}). We also track the values of the dynamic order parameter $\langle|Q| \rangle$ (see Fig.~\ref{fig:2}) and we find that the value of $\langle |Q|\rangle$ changes from zero to a non-zero number with increasing $p$. The system is therefore sensitive to NNN interactions. This sensitivity can be physically explained by considering the effect of interactions on the metastable lifetime. 
\subsection{Metastable lifetime, $\tau(h_{0},T,L,p)$ \label{sec:tau}} 
\begin{figure}[htb!]
\centering
{\includegraphics[width=3.5in]{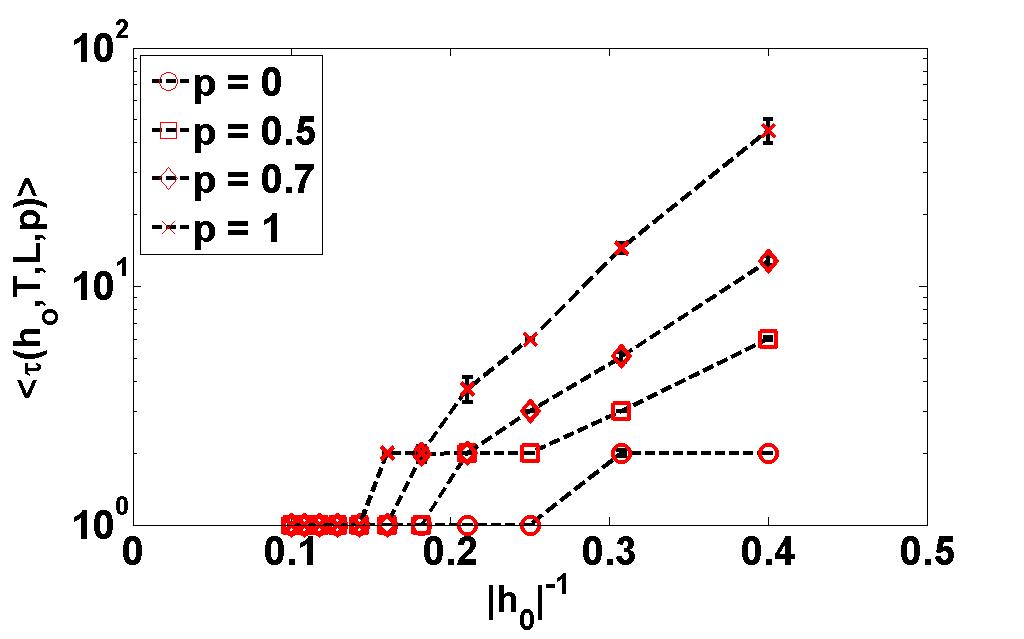}}
\caption{Variation of the average metastable lifetime, $\langle\tau(h_{0},T,L,p)\rangle$, as a function of interaction strength ratio $p=J_{nnn}/J_{nn}$ on a log-linear scale. The average lifetime was computed using the first passage of time to zero magnetization for 1000 repeated trials for different values of the external magnetic field amplitude $h_{o}$. The calculations were performed for $L$=128 at $T=0.8T_{c}^{NN}$. We see from the plot that the value of $\langle\tau\rangle$ is sensitive to additional interactions present in the system. The computed error bars (standard deviation) are displayed. For some of the computed data points the error bars are smaller than the symbol size.}\label{fig:3}
\end{figure}

The key to understanding the observed phenomena in the NNNKI model is to study the metastable lifetime. To determine $\tau$ for this model we performed several instantaneous field reveral simulations ~\cite{Rikvold:PhysRevE.49.5080}. The system was initially prepared in an all up spin configuration. The field was instantaneously reversed and the relaxation of the system then studied. The metastable lifetime is the number of MCSS needed for the system to decay to a net zero magnetized state ($m(t)=0$) from a completely magnetized state ($m(t)=\pm1$). The average metastable lifetime, $\left\langle \tau(h_{o},T,L,p)\right\rangle$, is calculated after 1000 repeated trials. The simulations were performed for $L$=128 at $T=0.8T^{NN}_{c}$ for various ratios of the interaction strength $p$=0, 0.5, 0.7, and 1. The results are shown in Fig.~\ref{fig:3}. From Fig.~\ref{fig:3} we see that the NNNKI model's average $\left\langle \tau(h_{o},T,L,p)\right\rangle$ is much greater than the NNKI model's $\left\langle \tau(h_{o},T,L,p)\right\rangle$. The lifetime is sensitive to the ratio of the interaction strengths. Since the metastable lifetime dictates the underlying metastable mode decay this in turn causes the system to exhibit a transition from the deterministic to the stochastic region. 
\section{Discussion and Conclusion}\label{sec:disconcl}
The computations in this article demonstrate that interactions have an important role to play in the kinetic Ising model in an oscillating applied field. While we have established that the system undergoes a transition from a deterministic to a stochastic regime our computations do not highlight the DPT or SR possible in the system. Some of the magnetization time series data ($p$ = 0.22 to 0.25) show an inkling of a MD region as seen in the ``wandering" time series graphs. However, the response of the system eventually becomes stochastic.  Our choice of a small lattice size ($L$ = 128) and system parameters ($L, h_{o}$, and T) may have caused the MD region to be less prominent compared to SR since the SR phenomenon is a finite-size effect. Investigating the model further for a DPT and/or SR for changing interaction strengths will be a topic of future investigation to be reported elsewhere. Another important conclusion of this article is that the metastable lifetime, $\tau$, which dictates the physics of the kinetic Ising model is sensitive to additional interactions. Finally, it is our hope that this work will motivate experimentalists to investigate the interaction induced transition observed in the NNNKI model in real material systems.
\begin{acknowledgements}
TD and WDB thank Per Arne Rikvold and Mark A. Novotny for many helpful discussions. WDB thanks Tom Colbert, Andy Hauger, and Pamplin Student Research Funds (ASU). TD thanks Faculty Research Faculty Development Funds (ASU).   
\end{acknowledgements}
\bibliography{text}

\begin{thebibliography}{15}
\expandafter\ifx\csname natexlab\endcsname\relax\def\natexlab#1{#1}\fi
\expandafter\ifx\csname bibnamefont\endcsname\relax
  \def\bibnamefont#1{#1}\fi
\expandafter\ifx\csname bibfnamefont\endcsname\relax
  \def\bibfnamefont#1{#1}\fi
\expandafter\ifx\csname citenamefont\endcsname\relax
  \def\citenamefont#1{#1}\fi
\expandafter\ifx\csname url\endcsname\relax
  \def\url#1{\texttt{#1}}\fi
\expandafter\ifx\csname urlprefix\endcsname\relax\def\urlprefix{URL }\fi
\providecommand{\bibinfo}[2]{#2}
\providecommand{\eprint}[2][]{\url{#2}}

\bibitem[{\citenamefont{Chakrabarti and
  Acharyya}(1999)}]{Chakrabarti:RevModPhys.71.847}
\bibinfo{author}{\bibfnamefont{B.~K.} \bibnamefont{Chakrabarti}}
  \bibnamefont{and} \bibinfo{author}{\bibfnamefont{M.}~\bibnamefont{Acharyya}},
  \bibinfo{journal}{Rev. Mod. Phys.} \textbf{\bibinfo{volume}{71}},
  \bibinfo{pages}{847} (\bibinfo{year}{1999}).

\bibitem[{\citenamefont{Sides et~al.}(1998{\natexlab{a}})\citenamefont{Sides,
  Rikvold, and Novotny}}]{Sides:PhysRevLett.81.834}
\bibinfo{author}{\bibfnamefont{S.~W.} \bibnamefont{Sides}},
  \bibinfo{author}{\bibfnamefont{P.~A.} \bibnamefont{Rikvold}},
  \bibnamefont{and} \bibinfo{author}{\bibfnamefont{M.~A.}
  \bibnamefont{Novotny}}, \bibinfo{journal}{Phys. Rev. Lett.}
  \textbf{\bibinfo{volume}{81}}, \bibinfo{pages}{834}
  (\bibinfo{year}{1998}{\natexlab{a}}).

\bibitem[{\citenamefont{Sides et~al.}(1999)\citenamefont{Sides, Rikvold, and
  Novotny}}]{Sides:PhysRevE.59.2710}
\bibinfo{author}{\bibfnamefont{S.~W.} \bibnamefont{Sides}},
  \bibinfo{author}{\bibfnamefont{P.~A.} \bibnamefont{Rikvold}},
  \bibnamefont{and} \bibinfo{author}{\bibfnamefont{M.~A.}
  \bibnamefont{Novotny}}, \bibinfo{journal}{Phys. Rev. E}
  \textbf{\bibinfo{volume}{59}}, \bibinfo{pages}{2710} (\bibinfo{year}{1999}).

\bibitem[{\citenamefont{Korniss et~al.}(2000)\citenamefont{Korniss, White,
  Rikvold, and Novotny}}]{Korniss:PhysRevE.63.016120}
\bibinfo{author}{\bibfnamefont{G.}~\bibnamefont{Korniss}},
  \bibinfo{author}{\bibfnamefont{C.~J.} \bibnamefont{White}},
  \bibinfo{author}{\bibfnamefont{P.~A.} \bibnamefont{Rikvold}},
  \bibnamefont{and} \bibinfo{author}{\bibfnamefont{M.~A.}
  \bibnamefont{Novotny}}, \bibinfo{journal}{Phys. Rev. E}
  \textbf{\bibinfo{volume}{63}}, \bibinfo{pages}{016120}
  (\bibinfo{year}{2000}).

\bibitem[{\citenamefont{Korniss et~al.}(2002)\citenamefont{Korniss, Rikvold,
  and Novotny}}]{Korniss:PhysRevE.66.056127}
\bibinfo{author}{\bibfnamefont{G.}~\bibnamefont{Korniss}},
  \bibinfo{author}{\bibfnamefont{P.~A.} \bibnamefont{Rikvold}},
  \bibnamefont{and} \bibinfo{author}{\bibfnamefont{M.~A.}
  \bibnamefont{Novotny}}, \bibinfo{journal}{Phys. Rev. E}
  \textbf{\bibinfo{volume}{66}}, \bibinfo{pages}{056127}
  (\bibinfo{year}{2002}).

\bibitem[{\citenamefont{Sides et~al.}(1998{\natexlab{b}})\citenamefont{Sides,
  Rikvold, and Novotny}}]{Sides:PhysRevE.57.6512}
\bibinfo{author}{\bibfnamefont{S.~W.} \bibnamefont{Sides}},
  \bibinfo{author}{\bibfnamefont{P.~A.} \bibnamefont{Rikvold}},
  \bibnamefont{and} \bibinfo{author}{\bibfnamefont{M.~A.}
  \bibnamefont{Novotny}}, \bibinfo{journal}{Phys. Rev. E}
  \textbf{\bibinfo{volume}{57}}, \bibinfo{pages}{6512}
  (\bibinfo{year}{1998}{\natexlab{b}}).

\bibitem[{\citenamefont{Buend\'{\i}a and Rikvold}(2008)}]{buenda:051108}
\bibinfo{author}{\bibfnamefont{G.~M.} \bibnamefont{Buend\'{\i}a}}
  \bibnamefont{and} \bibinfo{author}{\bibfnamefont{P.~A.}
  \bibnamefont{Rikvold}}, \bibinfo{journal}{Phys. Rev. E}
  \textbf{\bibinfo{volume}{78}}, \bibinfo{pages}{051108}
  (\bibinfo{year}{2008}).

\bibitem[{\citenamefont{Park et~al.}(2004)\citenamefont{Park, Rikvold,
  Buend\'ia, and Novotny}}]{PhysRevLett.92.015701}
\bibinfo{author}{\bibfnamefont{K.}~\bibnamefont{Park}},
  \bibinfo{author}{\bibfnamefont{P.~A.} \bibnamefont{Rikvold}},
  \bibinfo{author}{\bibfnamefont{G.~M.} \bibnamefont{Buend\'ia}},
  \bibnamefont{and} \bibinfo{author}{\bibfnamefont{M.~A.}
  \bibnamefont{Novotny}}, \bibinfo{journal}{Phys. Rev. Lett.}
  \textbf{\bibinfo{volume}{92}}, \bibinfo{pages}{015701}
  (\bibinfo{year}{2004}).

\bibitem[{\citenamefont{Robb et~al.}(2008)\citenamefont{Robb, Xu, Hellwig,
  McCord, Berger, Novotny, and Rikvold}}]{robb134422}
\bibinfo{author}{\bibfnamefont{D.~T.} \bibnamefont{Robb}},
  \bibinfo{author}{\bibfnamefont{Y.~H.} \bibnamefont{Xu}},
  \bibinfo{author}{\bibfnamefont{O.}~\bibnamefont{Hellwig}},
  \bibinfo{author}{\bibfnamefont{J.}~\bibnamefont{McCord}},
  \bibinfo{author}{\bibfnamefont{A.}~\bibnamefont{Berger}},
  \bibinfo{author}{\bibfnamefont{M.~A.} \bibnamefont{Novotny}},
  \bibnamefont{and} \bibinfo{author}{\bibfnamefont{P.~A.}
  \bibnamefont{Rikvold}}, \bibinfo{journal}{Phys. Rev. B}
  \textbf{\bibinfo{volume}{78}}, \bibinfo{pages}{134422}
  (\bibinfo{year}{2008}).

\bibitem[{\citenamefont{Zhu et~al.}(2004)\citenamefont{Zhu, Dong, and
  Liu}}]{Liu:PhysRevB.70.132403}
\bibinfo{author}{\bibfnamefont{H.}~\bibnamefont{Zhu}},
  \bibinfo{author}{\bibfnamefont{S.}~\bibnamefont{Dong}}, \bibnamefont{and}
  \bibinfo{author}{\bibfnamefont{J.-M.} \bibnamefont{Liu}},
  \bibinfo{journal}{Phys. Rev. B} \textbf{\bibinfo{volume}{70}},
  \bibinfo{pages}{132403} (\bibinfo{year}{2004}).

\bibitem[{\citenamefont{He and Wang}(1993)}]{PhysRevLett.70.2336}
\bibinfo{author}{\bibfnamefont{Y.-L.} \bibnamefont{He}} \bibnamefont{and}
  \bibinfo{author}{\bibfnamefont{G.-C.} \bibnamefont{Wang}},
  \bibinfo{journal}{Phys. Rev. Lett.} \textbf{\bibinfo{volume}{70}},
  \bibinfo{pages}{2336} (\bibinfo{year}{1993}).

\bibitem[{\citenamefont{Choi et~al.}(1999)\citenamefont{Choi, Lee, Samad, and
  Bland}}]{PhysRevB.60.11906}
\bibinfo{author}{\bibfnamefont{B.~C.} \bibnamefont{Choi}},
  \bibinfo{author}{\bibfnamefont{W.~Y.} \bibnamefont{Lee}},
  \bibinfo{author}{\bibfnamefont{A.}~\bibnamefont{Samad}}, \bibnamefont{and}
  \bibinfo{author}{\bibfnamefont{J.~A.~C.} \bibnamefont{Bland}},
  \bibinfo{journal}{Phys. Rev. B} \textbf{\bibinfo{volume}{60}},
  \bibinfo{pages}{11906} (\bibinfo{year}{1999}).

\bibitem[{\citenamefont{Jiang et~al.}(1995)\citenamefont{Jiang, Yang, and
  Wang}}]{PhysRevB.52.14911}
\bibinfo{author}{\bibfnamefont{Q.}~\bibnamefont{Jiang}},
  \bibinfo{author}{\bibfnamefont{H.-N.} \bibnamefont{Yang}}, \bibnamefont{and}
  \bibinfo{author}{\bibfnamefont{G.-C.} \bibnamefont{Wang}},
  \bibinfo{journal}{Phys. Rev. B} \textbf{\bibinfo{volume}{52}},
  \bibinfo{pages}{14911} (\bibinfo{year}{1995}).

\bibitem[{\citenamefont{Landau and Binder}(2000)}]{LandauBinderbook}
\bibinfo{author}{\bibfnamefont{D.~P.} \bibnamefont{Landau}} \bibnamefont{and}
  \bibinfo{author}{\bibfnamefont{K.}~\bibnamefont{Binder}},
  \emph{\bibinfo{title}{A Guide to Monte Carlo Simulations in Statistical
  Physics}} (\bibinfo{publisher}{Cambridge University Press},
  \bibinfo{year}{2000}).

\bibitem[{\citenamefont{Rikvold et~al.}(1994)\citenamefont{Rikvold, Tomita,
  Miyashita, and Sides}}]{Rikvold:PhysRevE.49.5080}
\bibinfo{author}{\bibfnamefont{P.~A.} \bibnamefont{Rikvold}},
  \bibinfo{author}{\bibfnamefont{H.}~\bibnamefont{Tomita}},
  \bibinfo{author}{\bibfnamefont{S.}~\bibnamefont{Miyashita}},
  \bibnamefont{and} \bibinfo{author}{\bibfnamefont{S.~W.} \bibnamefont{Sides}},
  \bibinfo{journal}{Phys. Rev. E} \textbf{\bibinfo{volume}{49}},
  \bibinfo{pages}{5080} (\bibinfo{year}{1994}).

\end{thebibliography}
\end{document}